\documentstyle[prl,aps,epsfig,multicol,amssymb]{revtex}
\begin{document}
\draft

 \title{Crystallization and phase-separation in non-additive
binary hard-sphere mixtures} 
\author{A.A. Louis, R. Finken and
J.P. Hansen} \address{Department of Chemistry, Lensfield Rd,
Cambridge CB2 1EW, UK\\} \date{\today} \maketitle
\begin{abstract}
\noindent 
We calculate for the first time the full phase-diagram of an
asymmetric {\em non-additive } hard-sphere mixture.  The
non-additivity strongly affects the crystallization and the
fluid-fluid phase-separation. The global topology of the phase-diagram
is controlled by an effective size-ratio $y_{\text{eff}}$, while the
fluid-solid coexistence scales with the depth of the effective
potential well.
\end{abstract}

\begin{multicols}{2}
Entropically driven phase transitions have received much attention
lately because of their direct relevance to the observed
phase-behaviour of colloidal suspensions\cite{entropy}.  Theoretical
work has focussed on the asymmetric binary hard sphere (HS) system, a
deceptively simple mixture of large and small particles, which
exhibits an interesting competition between demixing into dilute and
concentrated  suspensions of large particles, driven by the
familiar osmotic depletion effect\cite{Asak54}, and freezing into an
ordered  crystalline phase.  Recent Monte Carlo (MC)
simulations\cite{Dijk98} of binary colloidal dispersions of spherical
particles with additive diameters $\sigma_1$ and $\sigma_2$, and size-ratio
$y= \sigma_2/\sigma_1 \leq 0.2$, have convincingly demonstrated that
the demixing transition conjectured earlier\cite{Bibe91} is always
preempted by a direct freezing of a low concentration disordered
(``fluid'') phase into an FCC crystal of large particles.  It was
argued elsewhere\cite{Bibe97} that a small degree of non-additivity of
the diameter $\sigma_{12}$, determining the distance of closest
approach between large and small particles, might drive the demixing
transition from metastable to stable.  This view has been supported by
recent MC simulations\cite{Dijk98b} which show that non-additivity
significantly lowers the packing fraction at the critical point of the
fluid-fluid coexistence curve.  However, to reach firm conclusions
concerning the observability of a fluid-fluid demixing transition, the
effect of non-additivity on the freezing transition must also be
considered explicitly, in order to map out a complete phase diagram.
This is precisely the objective of this letter, where the important
influence of non-additivity on the global phase behaviour of highly
asymmetric hard sphere  mixtures is evaluated within a systematic
Statistical Mechanical perturbation treatment.  Even a small
non-additivity is shown to have a large effect on the interpretation of
the phase behaviour, and experiments on sterically or
charge-stabilized binary ``HS'' colloids are shown to  generically
exhibit non-additive behaviour.

Consider a binary system of HS with distances of closest approach
$\sigma_{\alpha \beta}\, (1 \leq \alpha, \beta \leq 2)$, such that
$\sigma_{11} = \sigma_1$, $\sigma_{22} = \sigma_2$, and:
\begin{equation}\label{eq1}
\sigma_{12} = \frac{1}{2} ( \sigma_{11} + \sigma_{22} ) ( 1 + \Delta),
\end{equation}
where the non-additivity parameter $\Delta$ can be positive or
negative.  The case $\Delta =0$ corresponds to additive HS, and has
been widely studied with the usual techniques of the Statistical
Mechanics of fluids.  An extreme example of non-additivity is provided
by the case $\sigma_{22} =0$, $\Delta > 0$, which is the familiar
Asakura-Oosawa model\cite{Asak54} of colloid-polymer mixtures,
allowing for full interpenetrability of polymer coils.  This system is
predicted to exhibit phase separation into dilute and concentrated
colloid fluid phases for large enough $\Delta$\cite{Gast83,Lekk92}, in
good agreement with experimental findings\cite{Ilet95}.  The present
work is concerned with the case of small positive non-additivity for
HS mixtures with small size-ratios which would lead to a metastable
demixing transition in the additive limit, $\Delta = 0$. $\Delta > 0$
will obviously favour demixing since phases involving a majority of
particles of the same species will allow a more efficient
packing\cite{Fren92}.

To examine the possibility of a fluid-fluid phase separation for
$\Delta > 0$, it seems natural to develop a perturbation theory
expansion in powers of $\Delta$ around a reference mixture of
identical composition, involving additive HS.  The free energy per
unit volume $f_0 = F_0/V$, of the latter is rather accurately given by
the semi-empirical equation of state proposed by Mansoori {\em et.\
al.}\cite{Mans71}, which improves on the compressibility equation of
state derived from the analytic solution of the PY
equations\cite{Lebo64}.  Note that neither predicts a spinodal
instability for any size-ratio or composition, as characterized by a
vanishing determinant of the stability matrix $M = | \partial^2 f_0 /
\partial n_\alpha \partial n_\beta |$, where $n_\alpha = N_\alpha/V$
is the number density of $\alpha$ spheres. However for fixed packing
fraction, they predict that $M$ approaches zero as $y^3$\cite{Lebo64},
suggesting that for larger and larger size-asymmetry, smaller and
smaller perturbations to the purely additive case can drive the
stability matrix negative.  The additive case is in a sense
``marginal'' to phase-separation, which explains why the very
existence of such a separation is so sensitive to the particular
approximations used\cite{Bibe91}.

A direct application of standard thermodynamic perturbation
theory\cite{Hans86} to the unlike pair potential: $v_{12}^{(\lambda)}
(r) = v_{12}^{(0)} (r/(1 + \lambda \Delta))$, where
$\lambda = 0$ and $1$ correspond, respectively, to the additive
reference system with diameter $\sigma_{12}^{(0)}$, and to the
non-additive mixture of interest, leads to the first order correction
to the free energy:
\begin{equation}\label{eq3}
\frac{ \beta (F - F_0)}{V} = 4 \pi \Delta n_1 n_2 \sigma_{12}^{(0)\,3}
g_{12}^{(0)}(\sigma_{12}^{(0)}).
\end{equation}
In Eq.\ \ref{eq3}, $g_{12}^{(0)}(\sigma_{12}^{(0)})$ is the contact
value of the unlike pair distribution function of the reference
mixture taken from the analytic solution of the PY equation for an
additive binary HS mixture\cite{Lebo64}.
\vglue-0.2cm
\begin{figure}
\begin{center}
\epsfig{figure=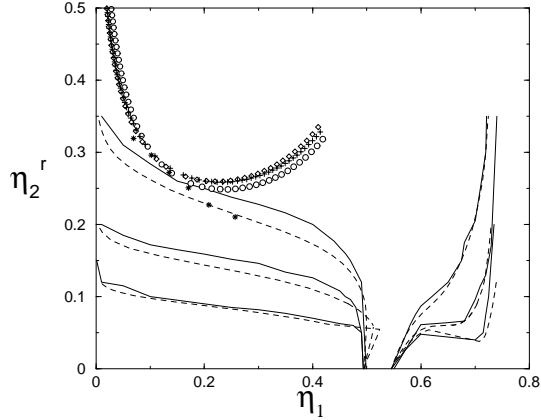,width=7cm} 
\begin{minipage}{8cm}
\caption{\label{Fig1}
The upper curves are fluid-fluid spinodals for $y=0.2, \Delta=0.033$
 as obtained from  the 4 theories described in the text:
Mansoori reference ($\circ$), PY reference ($\diamond$),
Barboy-Gelbart ($+$), and BPGG ($\ast$).  The lower 3 sets
of curves compare the fluid-solid binodals calculated by 1st order
perturbation theory in the potential of Eq.\ (\protect\ref{eq4})
(dashed lines) with MC results
\protect\cite{Dijk98}
 (solid
lines) for the additive case, $\Delta =0$.  From top to bottom, $y =
0.2, 0.1$, and $0.05$ respectively. 
}
\end{minipage}
\end{center}
\end{figure}
\vglue - 0.5cm
  The fluid-fluid spinodal
curves resulting from this two-component perturbation theory using
both the Mansoori\cite{Mans71} and PY\cite{Lebo64} reference $F_0$'s
are compared in Fig.\ \ref{Fig1} to an approximate series expansion
due to Barboy and Gelbart\cite{Bibe97,Barb79} for a size-ratio $y =
0.2$ and $\Delta = 0.033$.   
Note that the results are plotted for a semi-grand
ensemble: the binary mixture is in osmotic equilibrium with a
reservoir of small spheres which fixes their chemical potential
$\mu_2$; the thermodynamic variables controlling the phase behaviour
are the packing fractions, $\eta_1 = \pi n_1 \sigma_1^3/6$, of the
large spheres in the mixture and $\eta_2^r$ of small spheres in the
reservoir (or equivalently $\mu_2$).  In sharp contrast to the
additive case where different theories yield dissimilar results\cite{Dijk98}, all
3 approaches yield similar results for the spinodal curves. Moreover,
the are consistent with results from numerical solutions of the BPGG
integral equation\cite{Ball86}, at least for low values of $\eta_1$.  As
expected, the small correction due to the non-additivity triggers a
phase-separation which is absent in the two additive reference systems
$(\Delta =0$).  The demixing transition, which is marginal for
additive HS's\cite{Bibe91}, is strongly enhanced by a modest degree of
non-additivity.  Moreover, increasing $\Delta$ shifts the demixing
transition to lower and lower packing fraction $\eta_2^r$, of small
spheres, as shown in Fig.\ \ref{Fig2}.

The above two-component perturbation scheme cannot be adapted to
investigate the freezing of non-additive HS mixtures, mainly because
the crystal phase of the additive reference system is poorly
understood for small size-ratios $y$.  HS alloys form substitutionally
disordered crystals for $y \gtrsim 0.85$\cite{Barr86}, and interesting
super-lattice structures for $y \lesssim 0.6$\cite{Bart92}, but the
structure for smaller values of $y$ (say $y \lesssim 0.3$) is not well
understood, although it has been conjectured that the large spheres
might form an FCC lattice, permeated by a fluid of small spheres, at
least for sufficiently small $y$\cite{Cous98}.  To avoid these
difficulties, one may resort to an effective one-component
description, by integrating out the degrees of freedom of the small
spheres for any given configuration of large spheres.  This procedure
leads to effective interaction potentials between the large spheres
which are determined by the free energy (or grand potential) of the
inhomogeneous fluid of small spheres in the ``external field'' of the
larger particles. The {\em basic philosophy} behind effective
potentials in complex fluids is that the (considerable) initial effort
spent deriving accurate effective potentials is recouped when they are
used as input into the well developed machinery of liquid state
theory\cite{Hans86}. In the case of colloid-polymer mixtures, this
approach leads to the Asakura-Oosawa effective pair potential between
large spheres\cite{Asak58}, and for additive HS mixtures the effective
pair potential has recently been extended to include excluded volume
correlations between small spheres\cite{Mao95,Dijk98}.  The procedure
used in these references may be generalised to derive the following
effective pair potential between large spheres in a non-additive HS
mixture:
\begin{equation} \begin{array}{l}\label{eq4}
\beta V_{\text{eff}}(r) = \infty \,\,\,\, ; \,\, r \leq \sigma_1 \,
\nonumber \\ \beta V_{\text{eff}}(r) = \displaystyle{ \frac{-3
\eta_2^r (1+y_{\text{eff}})}{2 \, y^3}} \biggl\{h(r)^2+\eta_2^r
\left[4\,h(r)^2-3\,y \,h(r) \right]
\vspace*{5pt} \nonumber \\ + (\eta_2^r)^2 \left. \left[ 10\, h(r)^2 -
12\,y\,h(r) \right] \biggl\} \right. \,\,\,\,; \,\, \sigma_1
\leq r \leq \sigma_1(1 + y_{\text{eff}}),
\end{array}\end{equation}
where the effective size-ratio is:
\begin{equation}\label{eq5}
 y_{\text{eff}} = y + \Delta + \Delta \, y,
\end{equation}
while the function $h(r) = (1 + y_{\text{eff}}) - r/\sigma_1$.  The
term linear in $\eta_2^r$ is the purely attractive Derjaguin form of
the Asakura-Oosawa potential, with the effective size-ratio
$y_{\text{eff}}$, while the higher order terms describe the partially
repulsive effects of the correlation-induced layering of the small
spheres around the large spheres.  As shown in the inset of Fig.\
\ref{Fig2}, increasing $\Delta$ at fixed $\eta_2^r$ deepens the
attractive well, while the correlation-induced repulsive barrier
remains roughly the same.

The effective interaction for $r > \sigma_1$ is treated as a
perturbation of the one-component HS reference system, and standard
first order perturbation theory\cite{Hans86,Gast83} is applied to
calculate the free energy and determine the fluid-solid phase
coexistence.  The predictions of this effective one-component
perturbation theory are compared in Fig.\ \ref{Fig1} to MC data for
the additive ($\Delta =0$) case (which in turn compares well with full
2-component MC simulations\cite{Dijk98}); the agreement is seen to be
good, thus lending confidence to the extension of the theory to the
non-additive case, for which no MC data is available. Interestingly
this success is not mirrored for the fluid-fluid transition: the
1-component perturbation theory predicts a critical point at much
higher values of $\eta_1$ than was found in the simulations,
demonstrating that care must be taken when applying ideas culled from
the field of simple fluids to the effective potentials arising from
complex fluids.  On the other hand, the two-component perturbation
theory embodied in Eq.\ (\ref{eq3}) turns out to be much more accurate
and has been used throughout to determine the demixing of non-additive
HS's into two fluid phases.

Results for $0.02 \leq \Delta \leq 0.25$ are shown in Fig.\
\ref{Fig2}.  Both the fluid-fluid spinodal and the liquidus line of
the freezing transition are seen to shift to lower $\eta_2^r$ as
$\Delta$ increases, so that the fluid-fluid coexistence curve remains
metastable up to $\Delta \backsimeq 0.2$, at which stage the critical
point becomes stable.  The two-component perturbation theory cannot be
trusted for significantly larger values of $\Delta$, because of its
first order nature.  On the other hand non-additivity is expected to
remain rather small for sterically or electrostatically stabilized
colloids, so that the present calculations imply that it may be
unlikely to observe fluid-fluid phase coexistence in asymmetric
colloidal mixtures.
\vglue -0.3cm
\begin{figure}
\begin{center}
\epsfig{figure=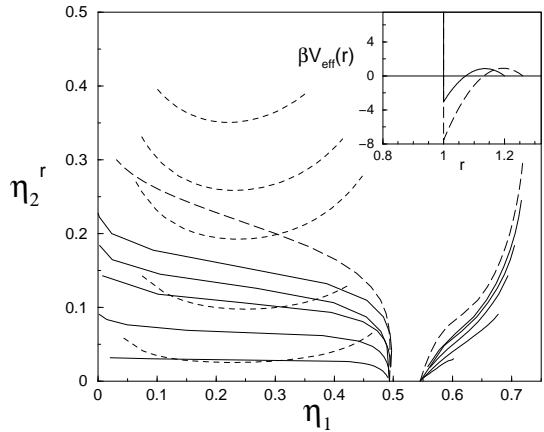,width=7cm}
\begin{minipage}{8cm}
\caption{\label{Fig2} Fluid-fluid spinodals (dashed lines) and
fluid-solid binodals (solid lines) for $y = 0.2$, and $\Delta = 0.02,
0.033, 0.05, 0.1, \& 0.25$ (from top to bottom). The long-dashed curve
is the fluid-solid binodal for $\Delta=0$.  Inset: effective
potentials of Eq.\ (\protect\ref{eq4}) for $y=0.2, \eta_2^r =0.3,
\Delta = 0$ (solid line) and $\Delta=0.05$ (dashed line)}
\end{minipage}
\end{center}
\end{figure}
\vglue - 0.5cm

\begin{figure}
\begin{center}
\epsfig{figure=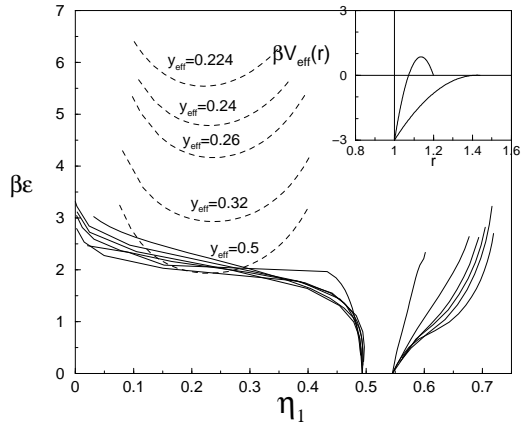,width=7cm} 
\vglue 0.1cm
\begin{minipage}{8cm}
\caption{\label{Fig3} Same as Fig.\ \protect\ref{Fig2} but with the
y-axis re-scaled w.r.t.\ the well-depth at contact of Eq.\
(\ref{eq4}). Inset: two typical effective potentials from Eq.\
(\ref{eq4}) for $y_{\text{eff}} = 0.2$, and $y_{\text{eff}} = 0.44$;
both result in very similar liquidus lines!  }
\end{minipage} 
\end{center}
\end{figure}

However, the calculations lead to another unexpected prediction.  If
the phase diagrams in Fig.\ \ref{Fig2} are rescaled, by choosing the
reduced well depth at contact $ \beta \epsilon = \beta V_{\text{eff}}
(\sigma_1)$, rather than $\eta_2^r$, as the thermodynamic variable
along the $y$ axis, the fluid-solid liquidus curves for all values of
$\Delta$ fall practically on top of each other.  This
``quasi-universal'' behaviour is shown in Fig.\ \ref{Fig3}: the
fluid-solid liquidus line scales with well-depth at contact, and is
largely independent of the shape or range (characterized by
$y_{\text{eff}}$) of the effective pair potential between the larger
spheres. 
\vglue -0.5cm
\vglue -0.5cm
This behaviour contrasts sharply with the fluid-fluid
coexistence curves, which are very sensitive to the range and shape of
the effective potential.  Note that the fluid-fluid demixing
transition becomes stable for $y_{\text{eff}} \gtrsim 0.4$, which is
comparable to the results of direct simulations of the Asakura-Oosawa
model for colloid-polymer mixtures\cite{Meij94}.

The possibility of phase-separation in binary hard-sphere mixtures
inspired a series of experiments on hard-sphere
like-colloids\cite{Sany92}.  The hard-sphere repulsion is achieved by
steric stabilization, with a repulsive outer brush of co-polymer, or
else the colloids are stabilized by repulsive electrostatic
interactions, with enough salt added to obtain a very short screening
length.
Interactions between charge-stabilized binary colloids are well
described by the DLVO potential\cite{Isra92,Croc96}.  For sterically
stabilized colloids, the interaction between two spheres can be
estimated by combining the interaction between 2 flat brush-covered
surfaces, which is well described by the Alexander--de Gennes
theory\cite{Isra92}, with the Derjaguin approximation, which results
in: $V_{22}(r-\sigma_{22}) = y \, V_{11}(r-\sigma_{11})$, and
$V_{12}(r-\sigma_{12}) = (2y/(1+y)) V_{11}(r-\sigma_{11})$, where
$V_{\alpha \beta}(r-\sigma_{\alpha \beta}))$ is the interaction
potential between the brushes of species~$\alpha$ and~$\beta$.  The
repulsive interaction is smaller between smaller particles because the
brush surface overlap area is smaller as a function of distance $r$.
The 3 effective hard-sphere diameters can be estimated by the diameter
at which the interaction is a few times $k_B T$, and then used to
determine the non-additivity.  For both sterically and
electrostatically stabilized binary colloidal mixtures this results
in:
\begin{equation}\label{eq6}
\Delta = \frac{l}{\sigma^{(0)}_{12}} \ln \left( \frac{2 \sqrt{y}}{1 +
 y} \right) + {\cal O}(\frac{l}{\sigma^{(0)}_{12}})^2,
\end{equation}
where $l$ is the decay length of the brush\cite{Isra92}, or the Debye
screening length of the DLVO interaction.  To first order, the
non-additivity depends neither on the pre-factors in the interactions,
nor on exactly how many times $k_BT$ is chosen as the hard-sphere
criterion; only the ``softness'' of the potential enters.  Since the
logarithmic factor in Eq.\ (\ref{eq6}) is negative, sterically or
electrostatically stabilized colloidal mixtures will manifest {\em
negative} non-additivity.  For example, the popular model colloids
made of PMMA (polymethylmethylacrylate) cores are stabilized by a PHS
(poly-12-hydroxysteric acid) brush, typically $10-12$ nm wide, with $l
\approx 3-4$ nm\cite{Isra92}, a number confirmed by direct
measurements with a surface-force apparatus\cite{Cost92}.  Thus a
mixture with the larger species at $400$ nm and the smaller at $40$ nm
(y=0.1) will have a negative non-additivity of about $\Delta = -0.01$;
i.e., the cross-diameter $\sigma_{12}$ is $1 \%$ less than the
additive cross-diameter $\sigma^{(0)}_{12}$. 
\vglue-0.5cm
\begin{figure}
\begin{center}
\epsfig{figure=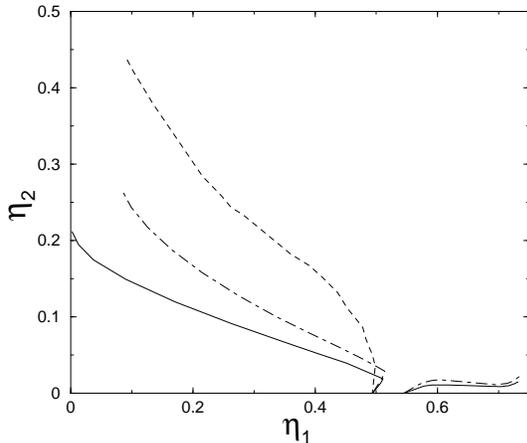,width=7cm}
\vglue -0.2cm
\begin{minipage}{8cm}
\caption{\label{Fig4}Fluid-solid binodals for $y = 0.1075$, $\Delta
= 0$ (solid lines) and $\Delta = -0.01$ (dash-dotted lines) plotted for
absolute $\eta_2$, (not $\eta_2^r$) to compare to the liquidus line
from Imhof and Dhont\protect\cite{Sany92} for the same $y$ (dashed
line).}
\end{minipage}
\end{center}
\end{figure}
\vglue - 0.5cm

Fig.\ \ref{Fig4} demonstrates that even a small amount
of negative non-additivity has a relatively large effect on the
phase-diagram, raising the liquidus line to higher effective values of
$\eta_2$. No fluid-fluid de-mixing is expected for $\Delta < 0$.

In conclusion then, we have demonstrated that non-additivity has an
important effect on the phase-diagram of asymmetric binary hard-sphere
mixtures, affecting both the fluid-fluid and the fluid-solid lines.
The fluid-solid liquidus lines show a near-universal behaviour, and
experimental hard-sphere like colloidal dispersions are expected never
to separate into two fluid phases.

We would like to thank R. Evans, M. Dijkstra, M. Child, Y. Mao,
P. Warren, W. Poon, P. Pusey, M. Cates, and D. Frenkel for helpful
discussions. AAL was supported by the EC through grant
EBRFMBICT972464.

\end{multicols}
\end{document}